\documentclass[hidelinks]{extarticle}
\usepackage{setspace}
\usepackage{authblk}
\usepackage{amsmath}
\usepackage[utf8]{inputenc}
\usepackage{multicol}
\usepackage{rotating}
\usepackage{hyperref}
\usepackage{float}
\usepackage[dvipsnames]{xcolor}
\usepackage{amssymb}
\usepackage{pifont}
\usepackage{booktabs}
\usepackage{graphicx}
\usepackage[toc,page]{appendix}

\usepackage[normalem]{ulem}
\useunder{\uline}{\ul}{}

\usepackage
[a4paper,
        left=2.5cm,
        right=2.5cm,
        top=2.5cm,
        bottom=2.5cm,]
{geometry}

\usepackage{footmisc}

\usepackage[english]{babel}
\usepackage[authoryear]{natbib}
\usepackage{caption}
\graphicspath{{./figures/}}

\usepackage{tabularx} 
\usepackage{threeparttable}

\def\sym#1{\ifmmode^{#1}\else\(^{#1}\)\fi}

\usepackage{float}
\usepackage{booktabs}

\begin{document}

\title{There are different shades of green: heterogeneous environmental innovations and their effects on firm performance}
\date{\today}

\author[$\circ$]{Gianluca Biggi}
\author[$\circ$,$\S$]{Andrea Mina}
\author[$\circ$]{Federico Tamagni}

\affil[$\circ$]{\footnotesize{Institute of Economics, Scuola Superiore Sant'Anna, Pisa, Italy}}
\affil[$\S$]{Centre for Business Research, University of Cambridge, Cambridge, UK}

\maketitle 

\noindent \textbf{Abstract}:
Using a firm-level dataset from the Spanish Technological Innovation Panel (2003-2016), this study explores the characteristics of environmentally innovative firms and quantifies the effects of pursuing different types of environmental innovation strategies (resource-saving, pollution-reducing, and regulation-driven innovations) on sales, employment, and productivity dynamics. The results indicate, first, that environmental innovations tend to be highly correlated with firms' technological capabilities, although to varying degrees 
across types of environmental innovation, whereas structural characteristics are less significant. Second, we observe heterogeneous effects of different types of environmental innovation on performance outcomes. We find no evidence that any type of environmental innovation fosters sales growth while pollution-reducing and regulation-driven innovations boost employment growth. Moreover, both resource-saving and pollution-reducing innovations bring about productivity advantages.

\bigskip


\newpage

\section{Introduction}
\label{sec:intro}
As concerns about climate change, resource scarcity and pollution continue to escalate, firms are under increasing pressure to integrate environmental objectives into their operations. Growing public scrutiny and the implementation of stricter regulatory measures have induced firms to engage in innovation efforts that can produce positive - or less negative - environmental outcomes \citep{delmas2008organizational,horbach2022climate}. Such innovative efforts, variously referred to as environmental innovations,  green innovations, sustainable innovations, or eco-innovations, have gained significant policy and scholarly attention in recent years~\citep{ec2019communication}, as the transition towards environmentally conscious practices and the ensuing impact on economic performance inform a broad debate on the trade-offs between sustainability and 
economic growth~\citep{rennings2004effect,kolk2016social}. 

Although various notions of environmental innovation exist in the literature~\citep{boons2013business, ghisetti2014environmental, ziegler2018disentangling}, they converge in defining this kind of innovations as new solutions, methods and processes related to reducing environmental harm and promoting the efficient use of natural resources~\citep{hojnik2016drives}. While this definition theoretically encompass a broad spectrum of practices --ranging from environmental management practices to more strategic approaches integrating environmental and economic considerations through environmental innovation initiatives~\citep{porter1995toward}-- the empirical assessment of environmental innovations involves three types of innovations, related to reduction of materials and energy used in production (resource-saving innovations), reduction of negative environmental externalities (pollution-reducing innovations), and compliance with environmental regulations (regulation-driven innovations). 
More specifically, resource-saving practices involve the development and implementation of technologies, processes, or strategies aimed at optimizing resource or energy consumption, typically via improving resource efficiency or reducing waste. Pollution-reducing innovations target the adoption of technologies and practices that minimize the release of harmful emissions and pollutants into the environment. Innovations driven by regulatory compliance involves adherence to environmental regulations and standards set by authorities, ensuring that firms operate within legal frameworks \citep{popp2019environmental}.

These three types of environmental innovations represent very different "ways to be green", and each of them is likely to involve specific 
knowledge and actions. Existing studies tend to adopt a generic notion of what an eco-innovation is, and generally pay little or no attention to the heterogeneity that exists across various types of environmental innovations actually undertaken by firms. In this paper, we dig deep into this heterogeneity, addressing two research questions. First, we ask whether there are some distinctive characteristics across firms undertaking different types of environmental innovations, and what these characteristics might be. Second, we ask what effects the different types of environmental innovation have on different dimensions of firm performance. 

Despite the growing interest in environmental innovation, both questions are still awaiting conclusive answers in the literature. On the one hand, the lack of comprehensive understanding of the 'identity cards' of firms undertaking different types of environmental innovation represents a well-known gap among existing studies~\citep{rennings2000redefining, ziegler2018disentangling}. Uncovering firm characteristics associated with different eco-innovations shall provide a much more precise qualification of the drivers of environmental innovation \citep{de2012environmental, ghisetti2014environmental, marzucchi2017forms}. 
We address this gap by exploring how firms involved in the different types of environmental innovations compare, in terms of key demographic, financial and technological characteristics, such as innovative capabilities, firm size, financial resources, and industry. 

Conversely, while we have evidence on the effect of generic eco-innovation for firm performance, to the best of our knowledge, there is no study exploring in a systematic quantitative way the specific impact of different types of environmental innovations. This is a critical research gap, which needs to be addressed in order to obtain a more accurate assessment of the potential trade-offs and synergies between environmental and economic goals \citep{dangelico2010mainstreaming, ambec2008does,schiederig2012green}. To answer these questions, we adopt a multi-dimensional approach, taking three complementary measures of performance outcome, namely sales growth, employment growth and productivity, thus asking whether our focal types of environmental innovations pay off, respectively, in terms of market success, job-creation and efficiency. This approach allows us to consider the diversity of impacts that is often neglected or washed away in existing studies linking the adoption of environmental innovations to performance~\citep{barbieri2017survey}. 

The empirical analysis exploits data generated by a rich yearly survey of Spanish firms over the period 2003-2016. The panel data combine information on firm financials, innovation behavior and, crucially, the importance of resource-saving, pollution-reducing and compliance-related in the firm's innovation activities. To obtain clean identification of effects, we isolate firms that focus on one type of environmental innovations, but not others, and employ statistical techniques suitable for identifying causal effects. 

Our empirical results provide evidence of the heterogeneous nature and impact of environmental innovation at the firm level. We find no evidence that any type of environmental innovation fosters sales growth, while pollution-reducing and compliance-related innovations boost firm employment. In addition, both resource-saving and pollution-reducing innovations generate productivity advantages.

\section{Background literature}
\label{sec:literature}
In this section we briefly review the literature related to our two main research questions. In general, existing studies tend to apply a generic definition of environmental innovation and of eco-innovating firms. For instance, starting from patenting activity of firms in specific technological areas classified as "green", or from identification of "green practices" via surveys and other sources eliciting generic environmental concerns from directors or CEO. While all these represent interesting proxies of firms' involvement in developing environmental-friendly new products or processes, none of the existing studies --to the best of our knowledge at least-- examines separately the three types of environmental innovations, as we do here.

\subsection{Environmental innovations and firm characteristics}
When searching for the drivers that shape the adoption of generically defined environmental innovations, existing studies have identified three main groups of factors: market-pull, technology-push, and policy-pull factors~\citep{horbach2012determinants, frondel2007end, guoyou2013stakeholders, kesidou2012drivers}. 

Market-pull factors are primarily driven by demand and preferences of consumers. As customers become more environmentally conscious and their preferences shift towards sustainable products and services, firms increasingly recognize that embracing environmental innovation can lead to competitive advantages, increased market share, and enhanced long-term profitability. This encourages firms to invest in innovations that reduce their environmental footprint, such as using renewable materials, reducing energy consumption, and minimizing pollution and waste generation \citep{eccles2014impact}. For instance, \cite{horbach2008determinants} documents that the expectations of new sales related to meeting green demand by consumers is key for environmental innovation in the context of German manufacturing. \cite{wagner2007relationship} underlines how collaboration with predominantly environmentally concerned stakeholders, such as active consumer associations, plays an essential role in the development of environmental innovations by firms. More generally, demand-pull factors have been shown to play a crucial role in driving investment in environmental innovation by firms \citep{triguero2013drivers,liao2019innovation}. The demand from the public sector also matter, as we observe an increasing demand for green-products and processes by the government, witnessed in recent years, endorsing environmental innovations through public procurement \citep{ghisetti2017demand}. 

Technology-push factors highlight the importance of technological change and knowledge creation in stimulating the adoption of environmental innovations. Technological advancements in renewable energy, materials science and manufacturing processes, create new opportunities for firms to invest in new technologies that enhance operational efficiency, reduce environmental impact and improve competitiveness~\citep{alkhuzaim2021evaluating}. As these new technologies emerge, they open up new avenues for developing and/or adopting more efficient and sustainable production processes, waste reduction techniques, and cleaner energy sources~\citep{horbach2022climate}. For instance, automation, data analytics, and Internet of Things devices help monitor and manage energy consumption, water usage, and other resources, leading to increased efficiency and reduced environmental impact. As firms' investment in R\&D to discover and/or apply novel technologies that can solve specific environmental challenges, technology-push may also lead to the creation of new products (good or services) and new processes with a lower environmental impact. Often, technology-push factors unfold through collaborations between firms, research institutions, and governmental bodies. Partnerships can facilitate the exchange of knowledge and resources, accelerating the development and deployment of new green technologies~\citep{de2022environmental}. 

Policy-pull factors encompass a vast array of regulations (ambient standards, technology-based standards, and performance-based standards) and market-based instruments (Pigouvian taxes, subsidies, deposit/refund systems, and tradable permits) that aim at preventing or reducing firms’ pollution activities. Their role in stimulating the adoption of environmental innovations by firms has received vast scholarly attention~\citep{brunnermeier2003determinants,rennings2011impact}. The empirical firm-level studies suggest that environmental regulations are positively associated with the probability that firms engage in environmental innovation~\citep{cleff1999determinants, frondel2007end}. In fact, regulations trigger investment in environmental R\&D, as this is seen by firms as a way to cut the costs of complying with environmental regulations \citep{brunnermeier2003determinants, horbach2008determinants, mazzanti2006economic, popp2006international}. In turn, regulations also stimulate the creation of or entry into new markets for eco-products. Indeed, regulations may be the only means to break existing technological lock-ins and move towards environmental-friendly technologies which usually have higher costs, particularly in the short term~\citep{klaassen2005impact}.

The study of the distinctive firm-level characteristics that firms should possess to be identified as environmental innovators, is to some extent less developed. Existing works have explored attributes such as size, industry, geographical location, technological capabilities, financial resources and knowledge sourcing. Results suggest that eco-innovators exhibit specific traits. They tend to exploit agglomeration economies~\citep{cainelli2015does}, leverage external sources of knowledge from suppliers, users, or public scientific organizations~\citep{de2013knowledge, ghisetti2015open}, invest in human capital endowment and employees training~\citep{horbach2008determinants, ghisetti2015investigating}.

In the empirical analysis, we expand the set of firm-attributes considered, and examine their possibly heterogeneous role as determinants of firms' involvement in resource-saving, pollution-reducing and regulation-driven innovations.

\subsection{Environmental innovation and firm performance}
Concerning our second research question, a small but growing literature asks how eco-innovating firms compare to other firms, in terms of various measures of firm performance, including the outcomes that we analyse --namely sales growth, employment growth, and productivity \citep{dangelico2010mainstreaming, horbach2013environmental, aldieri2021environmental}. However, there is a lack of comprehensive analyses considering the heterogeneity of these impacts. On the one hand, most studies take a generic definition of environmental innovation, overlooking the possibly nuanced effects of resource-saving, pollution-reducing, and regulation-driven innovations. On the other hand, there is a tendency to focus on one single dimension of firm performance, thus overlooking that different types of environmental innovations may have heterogeneous relations with alternative dimensions of firm performance~\citep{van2017revisiting, zheng2022friends}.

Resource-saving environmental innovations, by definition, enhance cost-efficiency by reducing energy or material consumption, and by optimizing production processes~\citep{aldieri2022research}. Their primary effect, therefore, is likely to improve productivity by reducing inefficiencies~\citep{mazzanti2009environmental}. However, they also have the potential to have a positive effect on sales growth, indirectly at least, to the extent that improved resource utilization translates into increased demand via more competitive pricing~\citep{schiliro2019sustainability} or quality improvement~\citep{herath2019impact}. In terms of employment growth, the effect is uncertain. One argument could be that firms may experience job losses following adoption of new technologies or operation restructuring related to resource-saving innovations~\citep{pfeiffer2001employment}. Conversely, resource-saving innovations may foster employment, if new dedicated personnel is hired to implement and manage them or if the ensuing competitive advantages are strong enough to foster expansion of operations and employment~\citep{zhou2021closing}. 

Pollution-reducing environmental innovations are likely to positively impact sales growth, as they contribute to enhance a firm's reputation and brand image by demonstrating a commitment to environmental sustainability, thus attracting environmentally conscious consumers ~\citep{delmas2018organizational}. However, this effect relies on consumer awareness and willingness to pay a premium for environmentally friendly products. Also, the successful adoption of these innovations can create new employment opportunities, and thus foster employment growth, particularly in the development, production, and maintenance of cleaner technologies~\citep{horbach2012determinants, kitzmueller2012economic}. Pollution reduction innovations can also have a positive effect on productivity, at least indirectly, if they involve optimizing production processes, reducing waste and improving resource efficiency. However, the actual impact on productivity varies depending on the specific characteristics of the innovation, industry context, and firm-specific factors~\citep{delmas2008organizational, melnyk2003assessing}.

The relationship between regulation-driven innovations and performance is perhaps more nuanced, as the actual impact heavily depends on the nature of the compliance strategy. On the one hand, as these innovations involve adapting business practices to meet regulatory requirements and environmental standards, they typically entail net costs for the firms which may offset potential benefits, thus potentially reducing efficiency and with no effect on sales and employment growth~\citep{dangelico2010mainstreaming}. However, indirect effects may arise, impacting on all the focal performance variables we consider. In terms of productivity, regulation-driven innovations can lead to process improvements, improved resource utilization and reduced waste, thus positively impacting efficiency~\citep{reyes2016corporate, managi2005environmental}. Also, innovations which show a proactive approach toward compliance with environmental regulations can enhance the firm's reputation, brand image and trust with environmentally conscious customers, thereby contributing to sales growth~\citep{cheng2022impact, simpson2004environmental}. Moreover, employment growth opportunities may arise if firms require new staff to master, adapt and monitor systems for environmental regulation compliance~\citep{porter1995toward, shrivastava1995role}. 

By separately estimating how firms involved in each type of environmental innovation compare to other firms in terms of the different performance dimensions, our analysis shed light on the diversity of potential impacts, often neglected in existing studies.

\section{Data and main variables}
\label{sec:data}
\subsection{Source and sample}
Our empirical analysis is based on a firm-level dataset drawn from the Spanish Technological Innovation Panel (henceforth PITEC). This is a well-known dataset, widely used in innovation studies, collected following the guidelines and practices of the Community Innovation Survey (CIS) and other standard innovation surveys around the world, according to the Oslo Manual \citep{oecd2005oslo}. The distinguishing feature of the PITEC is its longitudinal nature. Indeed, while CIS and similar international innovation surveys are run in waves of 2-3 years and do not follow the same set of firms across the different waves, PITEC data have a panel structure, thanks to systematic data collection which ensures, starting from 2003, consistent representativeness of the population of Spanish manufacturing and service firms over time, allowing to monitor the same firms over a good number of years. The data available for this study span the period 2003–2016, for a total of about 8,958 firms. The large majority of firms (57.70\%) are observed over the entire sample period. A further 20\% are present in the data for 10 years, while only a small percentage (7.41\%) are present for less than 5 years.

\subsection{Identifying environmental innovators}
In order to identify firms' involvement in environmental innovation, we start from specific questions of the PITEC survey, asking about the degree of importance which firms attach to specific green objectives in driving their innovation activity. The underlying assumption, common to other studies employing this kind of survey data, is that the declared level of importance is a good indicator to distinguish between firms that truly prioritize and actively engage in a given type of environmental innovation, vis-a-vis firms which have a low commitment or do not perform that type of environmental innovation at all~\citep{marzucchi2017forms, de2012environmental}.

The following items of the survey refer to the three types of environmental innovations we aim to analyse: 1) “Objectives of the technological innovation: Less materials per output” (OBJET9 - Importancia objetivo inn. tec.: Menos materiales por unidad producida); 2) “Objectives of the technological innovation: Less energy per output” (OBJET10 - Importancia objetivo inn. tec.: Menos energía por unidad producida); 3) “Objectives of the technological innovation: Reduced environmental impact” (OBJET11 - Importancia objetivo inn. tec.: Menor impacto medioambiental); 4) “Objectives of the technological innovation: Compliance with environmental, health or safety regulatory requirements” (OBJET13 - Importancia objetivo inn. tec.: Cumplimiento de los requisitos normativos medioambientales, de salud o seguridad). Firms answer ranking on a 1-4 scale the relevance of each type of environmental innovation objective for their realised innovation, where 1 = High importance,
2 = Medium importance, 3 = Low importance, 4 = Not relevant/not used. 

By considering the responses to these questions, we identify three groups of firms. First, we define as firms engaging in resource-saving innovations all those firms which assign high or medium importance to either using less materials or using less energy, while giving low or no importance to the other green objectives. Second, we define as  engaged in pollution-reducing innovation a firm which assigns high or medium importance to reducing environmental impact, but low or no relevance to the other objectives. Third, we identify as firms undertaking regulation-driven innovations the firms answering that regulation compliance has high or medium importance, while the other objectives have low or no importance. These definitions are meant to isolate engagement in each type of environmental innovation from the others, by considering as belonging to each given group only the firms having a focused commitment to the specific type of environmental innovation under consideration, while excluding firms that are only marginally or not at all engaged in the focal environmental innovation. We then map these groups into three dummy variables for "resource-saving", "pollution-reducing" and "regulation-driven" environmental innovation, respectively taking value 1 if a firm's answers meet the above definitions, in a given year. The reference counterfactual (the zeroes of the dummies) is defined by firms which assign low importance to or declare not to have engaged in \emph{any} of the three environmental innovations, in a given survey year. This eases identification, by avoiding to include in the control group firms with potentially high involvement in alternative environmental innovation strategies, when considering a given focal environmental innovation.\footnote{One could think of an even stricter definition, assigning to each group only the firms ranking a given environmental innovation objective as "highly important" (i.e., excluding "medium importance"). In preliminary analysis, we verified that such alternative identification would have left us with a too small number of treated units, however.}

\subsection{Performance outcomes}
Our analysis of the impact of different types of environmental innovation on firm outcomes, considers three complementary measures of performance. Sales growth is a measure of the ability of firms to translate environmental innovation into market success. Employment growth allows to evaluate the implications of environmental innovations for job-creation. Productivity is a proxy of the efficiency gains possibly arising from environmental innovation. The two growth variables are defined as log-differences over two consecutive years
\begin{equation} \label{1}
GrSales_{it}=\ln(S_{it})- \ln(S_{it-1})
\end{equation}
and 
\begin{equation} \label{2}
GrEmployees_{it}=\ln(E_{it}) - \ln(E_{it-1})
\end{equation}
\noindent
where $S_{it}$ and $E_{it}$ are total sales and number of employees of firm $i$ in year $t$, respectively.
\noindent Regarding productivity, we take a standard labour productivity index (in logs), defined as 
\begin{equation} \label{3}
Productivity_{it}=log(S_{it}/E_{it}) \quad.
\end{equation}
Basic descriptive statistics in Table~\ref{tab:descr} show that "environmental innovators" tend to outperform, on average, along all the dimensions of performance. Differences in means are generally larger for firms engaging in regulation-driven innovations, particularly in terms of sales growth.

\begin{table}[thp]
\caption{Descriptive statistics, by environmental innovation types}
\resizebox{\textwidth}{!}{
\begin{tabular}{@{}llcccc@{}}
\toprule
\multicolumn{1}{c}{\textit{}} & \multicolumn{1}{c}{\textit{}} & \multicolumn{1}{c}{\textbf{Resource-saving}} & \multicolumn{1}{c}{\textbf{Pollution-reducing}} & \multicolumn{1}{c}{\textbf{Regulation-driven}} & \multicolumn{1}{c}{\textbf{Controls}} \\
\textit{} & \textit{} & (N = 3,386) & (N = 1,585) & (N = 2,918) & (N = 8,901) \\ 
\midrule
\textbf{Outcome variables}: &&&&&\\
\textit{GR\_Sales}        & Mean     & -.001   & .002    & .038  & -.016  \\
\textit{}                 & Std. Dev & .574    & .75     & .706  & .697   \\
\textit{GR\_Empl}         & Mean     & -.017   & -.001   & .003  & -.022  \\
\textit{}                 & Std. Dev & .256    & .226    & .246  & .311   \\
\textit{Productivity}     & Mean     & 11.775  & 11.839  & 11.69 & 11.635 \\
\textit{}                 & Std. Dev & 1.04    & 1.134   & 1.127 & 1.129  \\
\hline
\textbf{Firm characteristics}: &&&&&\\
\textit{R\&D\_Int}        & Mean     & 7.756    & 8.945   & 8.826 & 6.442  \\
\textit{}                 & Std. Dev & 5.968    & 5.682   & 5.569 & 6.08   \\
\textit{Prod\_Inno}       & Mean     & .62      & .656    & .707  & .564   \\
\textit{}                 & Std. Dev & .485     & .475    & .455  & .496   \\
\textit{Proc\_Inno}       & Mean     & .741     & .602    & .666  & .56    \\
\textit{}                 & Std. Dev & .438     & .49     & .472  & .496   \\
\textit{Employees}        & Mean     & 4.132    & 4.148   & 4.105 & 3.974  \\
\textit{}                 & Std. Dev & 1.492    & 1.625   & 1.542 & 1.592  \\
\textit{Age}              & Mean     & 3.032    & 3.081   & 2.974 & 3      \\
\textit{}                 & Std. Dev & .7       & .703    & .776  & .723   \\
\textit{Ext\_Fin\_Constr} & Mean     & .631     & .62     & .607  & .566   \\
\textit{}                 & Std. Dev & .483     & .485    & .488  & .496   \\
\textit{Subsidy}          & Mean     & .186     & .302    & .229  & .177   \\
\textit{}                 & Std. Dev & .389     & .459    & .42   & .382   \\
\textit{Patents}          & Mean     & .116     & .16     & .149  & .083   \\
\textit{}                 & Std. Dev & .32      & .367    & .356  & .276   \\
\textit{Coop}             & Mean     & .355     & .475    & .4    & .281   \\
\textit{}                 & Std. Dev & .478     & .499    & .49   & .449   \\
\textit{Export}           & Mean     & .699     & .673    & .634  & .605   \\
\textit{}                 & Std. Dev & .459     & .469    & .482  & .489   \\
\textit{Group}            & Mean     & .44      & .428    & .38   & .381   \\
\textit{}                 & Std. Dev & .496     & .495    & .485  & .486   \\ 
\bottomrule
\end{tabular}
}
\label{tab:descr}
\end{table}

\subsection{Firm characteristics}
Much like other CIS-type innovation surveys, the PITEC combines an extremely rich source of information about a wide range of firms' innovation-related activities, with a somewhat more limited number of variables capturing firms' structure and industrial performance. We select a list of variables apt to providing a comprehensive picture of demographic, financial and technological characteristics. These variables are used as predictors of the probability to belong to each group of "environmental innovators", in tackling our first research question. They also serve as control variables to condition the estimation of the effect of environmental innovation on performance outcomes, responding to our second research question.

A first group of variables encompasses the standard set of industrial and structural characteristics typically available from innovation surveys. These include: a proxy of size in terms of the number of employees (\emph{Employees}); firm age (\emph{Age}) computed as years since foundation year; two dummy variables, respectively indicating a firm's affiliation to a business group~(\emph{Group}) and involvement in foreign export markets~(\emph{Export}); lagged values of sales growth, as a proxy for past opportunities; and a full-set of 2-digit sector dummies.

Among the available proxies of firm technological characteristics, we include the four measures of innovative inputs and outputs usually employed in innovation studies, namely intramural R\&D expenditures (\emph{R\&D\_int}), two dummies for product or process innovation (\emph{Prod\_Inno} and \emph{Proc\_Inno}), and a dummy for whether the firm filed at least one patent application in the focal year (\emph{Patents}). These attributes provide different ways to capture the overall capability of firms to undertake "generic innovation", which may obviously correlate with or spill-over into whether and how much firms may be able to "specialize" in environmental innovation. 

In addition, we also include a set of standard proxies for the way firms interact with the external environment in carrying out their innovative activity, helping to capture the contexts and circumstances that may influence firms' adoption of environmental innovation. This set of variables encompasses a dummy variable for whether the firm receives any form of public support to innovation (\emph{Subsidy}); a dummy variable indicating whether the firm encountered difficulties in accessing external finance for innovation purposes (\emph{Ext\_Fin\_Constr}); and a dummy indicating if the firm has formalised agreements to jointly develop innovation with other firms or institutions (\emph{Coop}). 

Basic descriptive statistics for firm characteristics, distinguishing between firms engaged in environmental innovations vs.~other firms, are reported in Table~\ref{tab:descr}. 

\section{Empirical analysis}

\subsection{Determinants of environmental innovation}
To address our first research question, we leverage the panel structure of the PITEC dataset and rely on a Random-Effect Probit to estimate the influence of firm attributes on the probability to carry out resource-saving, pollution-reducing or regulation-driven innovations. The dependent variables are the dummies for the three types of environmental innovation, as defined above. These dummies are potentially time-varying, since assignment to each group is re-evaluated on each yearly survey. As for the explanatory variables, the estimated models include 1-year lags of the firm characteristics described above, in order to partially tackle simultaneity, plus a full set of year fixed-effects, to control for aggregate and macro-level dynamics over time.


\begin{table}[thp]
\caption{Environmental innovation and firm characteristics - Random Effect Probit estimates}
\resizebox{\textwidth}{!}{%
  \begin{threeparttable}
\begin{tabular}{lccc}
\toprule
                             & Resource-saving & Pollution-reducing & Regulation-driven\\ \midrule
\textit{R\&D\_Int(t-1)}     & 0.0297***         & 0.0342***         & 0.0388***       \\
\textit{}                    & (0.00376)         & (0.00603)         & (0.00451)       \\
\textit{Prod\_Inno(t-1)}       & 0.125***          & 0.0995            & 0.234***        \\
\textit{}                    & (0.0423)          & (0.0632)          & (0.0516)        \\
\textit{Proc\_Inno(t-1)}       & 0.392***          & -0.0229           & 0.155***        \\
\textit{}                    & (0.0434)          & (0.0631)          & (0.0489)        \\
\textit{GR\_Sales(t-1)}        & 0.0183            & 0.0166            & 0.00345         \\
\textit{}                    & (0.0200)          & (0.0328)          & (0.0272)        \\
\textit{Employees(t-1)}        & 0.0322*           & 0.0221            & 0.0575**        \\
\textit{}                    & (0.0192)          & (0.0300)          & (0.0231)        \\
\textit{Age(t)}                 & -0.112            & 0.0857            & 0.0634          \\
\textit{}                    & (0.0932)          & (0.153)           & (0.112)         \\
\textit{Ext\_Fin\_Constr(t-1)} & 0.147***          & 0.134**           & 0.0176          \\
\textit{}                    & (0.0389)          & (0.0631)          & (0.0477)        \\
\textit{Subsidy(t-1)}          & -0.0214           & 0.252***          & -0.00187        \\
\textit{}                    & (0.0504)          & (0.0718)          & (0.0594)        \\
\textit{Patents(t-1)}          & 0.0460            & 0.263***          & 0.204***        \\
\textit{}                    & (0.0640)          & (0.0940)          & (0.0756)        \\
\textit{Coop(t-1)}      & 0.122***          & 0.287***          & 0.106**         \\
\textit{}                    & (0.0424)          & (0.0651)          & (0.0486)        \\
\textit{Export(t-1)}           & 0.111**           & -0.0523           & 0.0240          \\
\textit{}                    & (0.0486)          & (0.0747)          & (0.0574)        \\
\textit{Group(t-1)}            & 0.102**           & 0.103             & -0.0824         \\
                             & (0.0500)          & (0.0766)          & (0.0617)        \\
Constant                     & -2.477***         & -3.939***         & -3.078***       \\
                             & (0.326)           & (0.549)           & (0.395)         \\
Year FE              & YES               & YES               & YES             \\

Industry FE             & YES               & YES               & YES             \\
Observations                 & 23,186            & 20,715            & 21,689          \\
 \midrule
\end{tabular}%
    \begin{tablenotes}
      \small 
    \item {\it Notes:} Robust standard errors in
      parenthesis. Asterisks denote significance levels:
      \sym{*}\(p<0.1\), \sym{**}\(p<0.05\), \sym{***}\(p<0.01\).
    \end{tablenotes}
  \end{threeparttable}
}
\label{tab:probit_result}
\end{table}

Table~\ref{tab:probit_result} reports the results. Regarding the analysis of firms' engagement in resource-saving innovations, our findings support prior literature that emphasizes the role of R\&D investments and product innovation in enabling firms to offer environmentally friendly alternatives and capture market opportunities~\citep{nidumolu2009sustainability}. Furthermore, the positive and significant coefficient on process innovation suggests that optimizing production processes can lead to significant resource savings and improved efficiency~\citep{seuring2008literature}. The estimated positive relationship with external financial constraints suggests that financial limitations can trigger firms to seek innovative solutions that reduce costs via reduced resource consumption~\citep{hsu2021evaluating}. The positive association between cooperation in innovation and resource-saving environmental innovations emphasizes the collaborative nature of environmental innovation, where knowledge and resources are shared to implement resource-saving practices effectively~\citep{johannessen2001innovation}. Also, the positive and significant coefficients on \emph{Export} and \emph{Group} dummies suggest that international competition stimulate sustainable practices~\citep{shrivastava1995role} and that the latter are facilitated by sharing of resources and knowledge transfer mechanisms occurring within industrial groups~\citep{chen2012measuring}.

Regarding pollution-reducing environmental innovations, the proxies for technological activity of firms activity display overall a weaker role. In fact, we confirm the critical role of R\&D and patents in driving environmental sustainability~\citep{popp2006international,horbach2008determinants}, while product and process innovations are not statistically significant. Also in analogy with resource-saving environmental innovation, we find that external financial constraints and access to external knowledge via innovation collaborations, both stimulate development of innovative solutions related to reducing negative pollution externalities~\citep{zhou2021closing,johannessen2001innovation}. The positive and significant association with innovation subsidies is specific to pollution-reducing innovations, vis-a-vis the other types of environmental innovation, highlighting the role of government support in incentivizing and facilitating firms' engagement in pollution-reduction efforts~\citep{popp2010energy}. 

Lastly, the findings about the drivers of regulation-driven innovations confirm the importance of generic innovation capabilities --as proxied by R\&D, product and process innovation, and patents-- for the ability of firms to come up with innovations related to their activity of complying with environmental regulation~\citep{fernandez2018innovation,latan2018too,popp2010energy}. Moreover, our estimates also confirm the importance of innovation collaborations~\citep{johannessen2001innovation}.

Overall, our findings underscore the importance of considering a wide range of organizational, technological internal and external factors that can promote environmental innovation within firms. In general, innovation capabilities and other innovation-related variables tend to play a stronger role compared to industrial characteristics such as size, age, group affiliation or participation to export markets. Notwithstanding these common patterns, there is also a good deal of heterogeneity across the three different types of eco-innovations.

\subsection{Environmental innovation and firm performance}

To examine our second research question, we estimate via state-of-the-art IPWRA (Inverse Probability Weighting Regression Adjustment) methods~\citep{wooldridge2007inverse, sloczynski2022doubly} the Average Treatment Effect (ATE) of being active in the three types of environmental innovation, taking as outcome variables sales growth, employment growth and (labour) productivity. The IPWRA estimates employ weighted regression coefficients to compute averages of treatment-level predicted outcomes, with the weights being the inverse of the probabilities of treatment (p-scores). These are obtained from a preliminary Probit taking the three dummies for environmental innovation status as the dependent variable, and the same set of explanatory variables employed in RE-Probit analysis of the previous section. The ATE is then computed contrasting the averages of predicted outcomes across treated and untreated units. While Inverse Probability Weighting (IPW) may be sensitive to the correct specification of the propensity scores and Regression Adjustment (RA) heavily relies on the assumption that the relation between outcome and propensity scores is correctly specified, IPWRA is doubly-robust and allows to explicitly check for covariate balance.\footnote{See Appendix~A reports a standard battery of diagnostics supporting reliability of the IPWRA estimates.}

\begin{table}[thp]
\caption{Estimated average treatment effects of environmental innovation on performance}
\resizebox{\textwidth}{!}{%
      \begin{threeparttable}
\begin{tabular}{@{}lllll@{}}
\toprule
 &  & \multicolumn{1}{c}{Growth of Sales} & \multicolumn{1}{c}{Growth of Employees} & \multicolumn{1}{c}{Productivity} \\ \midrule
Resource-saving & ATE     & 0.0174973     & 0.0075782      & 0.056608**   \\
                            &         & (0.0130477)   & (0.0055065)    & (0.0238057)  \\
Pollution-reducing & ATE     & -0.024647     & 0.0194297**    & 0.0687103*   \\
                            &         & (0.0298223)   & (0.0077491)    & (0 .0415853) \\
Regulation-driven  & ATE     & 0.0204088     & 0.0224737***   & 0.0143148    \\
                            &         & (0.013405)    & (0.0060779)    & (0.0277478)  \\
                            \midrule
\end{tabular}
    \begin{tablenotes}
      \small 
    \item {\it Notes:} Robust standard errors in
      parenthesis. Asterisks denote significance levels:
      \sym{*}\(p<0.1\), \sym{**}\(p<0.05\), \sym{***}\(p<0.01\).
    \end{tablenotes}
  \end{threeparttable}
}
\label{tab:ipwra_result}
\end{table}

The estimated ATEs of being active in different types of environmental innovations are reported in Table~\ref{tab:ipwra_result}.\footnote{See Appendix~B for tables reporting the full set of coefficient estimates.} We observe interesting variation, both across the dimensions of performance and across the types of environmental innovations considered. First, we find no significant effect of environmental innovations on sales growth, irrespective of the type of environmental innovation considered. Several explanations may contribute to this negative result. On the one hand, environmental innovations often entail additional costs, such as related to specific R\&D expenses or to the acquisition of specialized equipment and workforce. These increased costs can lead to price premiums for eco-friendly products, which consumers may be reluctant to pay, particularly in price-sensitive markets \citep{tully2014people}. On the other hand, the return on green-investment in terms of sales growth may take longer to materialize, making it challenging to establish a significant relationship, particularly in the short term~\citep{hart2011invited}. From our results, these channels seem to completely off-set the possibly growth-enhancing effects due to improved brand image towards environmentally conscious consumers. 

Second, we find a positive effect of pollution-reducing and regulation-related innovations on employment growth. These types of innovations often necessitate the adoption of new technologies, processes, and practices, creating a demand for specialized skills and expertise. For instance, the implementation of renewable energy systems or waste management solutions may require hiring employees with technical knowledge in these areas~\citep{moreno2021exploring}. Firms engaging in pollution-reducing and regulation-driven innovations are likely to experience employment growth via these channels. 

Third, regarding the impacts on productivity, we find that firms engaging in resource-saving or pollution-reducing innovations enjoy productivity advantages. These results suggest that these two types of innovations, acting upon utilization of resources and reduced waste, help firms to improve efficiency, likely via streamlining of production processes. This in turn enables a more productive utilization of labour inputs, allowing employees to concentrate on value-adding activities rather than managing resource inefficiencies~\citep{seuring2008literature}. On the contrary, regulation-driven innovations do not exert any statistically significant effect on productivity.

Read together, our results underscore the complexity of the relationship between environmental innovations and firm performance, supporting the notion that there is no unique path from environmental innovation to superior performance. They also emphasize the importance of considering diverse dimensions of performance when evaluating the outcomes of environmental initiatives.

\section{Conclusions}
\label{sec:concl}
Our study provides evidence of the heterogeneous nature and impact of environmental innovations at the firm level. We highlight two significant contributions. 
First of all, we introduce a novel characterization of environmental innovation by studying similarities and differences across firms engaged in innovation activities driven by three distinct motives: resource saving, pollution reduction, and regulatory compliance. This categorization allows for a better understanding of environmental innovation strategies, improving upon the consideration, highly common in the existing literature, that environmental innovation is an all-encompassing and undifferentiated construct, rather than a more complex set of motivations and actions.

Second of all, we demonstrate that different dimensions of firm performance are influenced in different ways by environmental innovations. We have argued that there can be no homogenous 
expectation on all dimensions of firm performance as far as environmental innovation is concerned. Different environmental innovation strategies entail different growth prospects relative to specific performance outcomes. In our empirical analysis, we find no evidence that environmental innovations foster sales growth, as we find neither positive effects for pollution-reducing innovations (for example through reputation or signalling effects) nor negative effects for innovation related to regulatory compliance (via, for example, increased costs of production or firm organisational adjustments). Interestingly, pollution-reducing and compliance-driven innovations boost firm employment, and the most likely mechanism at play here is the need to acquire through external labour markets new skills that are required to manage these types of innovation investments. Finally, both resource-saving and pollution-reducing innovations bring about productivity advantages (this does not apply to compliance-driven innovation), an important finding in relation to the aims of profit-seeking firms and the presence of market incentives for environmental innovation.

More broadly, our study contributes to the debate on how the transition of firms towards environmental sustainability affects their economic performance. The new evidence we provide contains some indications that policies could and arguably should take into account the way in which different performance outcomes respond to different types of innovations. Bearing in mind that this is not a policy evaluation exercise, which would require a different methodological approach, we can observe that compliance-induced innovations do not seem to damage the performance of firms, which are instead able to adjust to new regulatory constraints. In addition, environmental policy designed to reduce pollution and meet compliance requirements could be combined with employment support objectives, at least in the absence of specific skills shortages. Higher education policies might indeed be required in cases of specific skills shortages. Moreover, the effect of environmental innovation does not appear to go through signalling mechanisms (there is no effect on sales) but productivity is enhanced by both pollution-reducing and resource-saving innovation. This is an interesting finding because these kinds of innovation appear to favour at the same time the competitive advantage of firms and their environmental sustainability, suggesting that there might be scope for green policy interventions that, by limiting negative environmental externalities or by reducing the use of resources, could simultaneously translate into market advantages. Further work could address potentially relevant boundary conditions for these effects (e.g. by testing the role of geographical and institutional contexts), and perhaps focus on the development within the firm of organisational capabilities that may make compatible with one another the achievement of economic returns and environmental sustainability targets.   

\newpage
\clearpage

\bibliography{main.1}

\newpage
\clearpage

\section*{Appendix~A: IPWRA diagnostics}

We here document about standard diagnostics supporting the reliability of the IPWRA estimates.

\begin{figure}[thp]
    \begin{center}
    \includegraphics[width=0.6\linewidth]{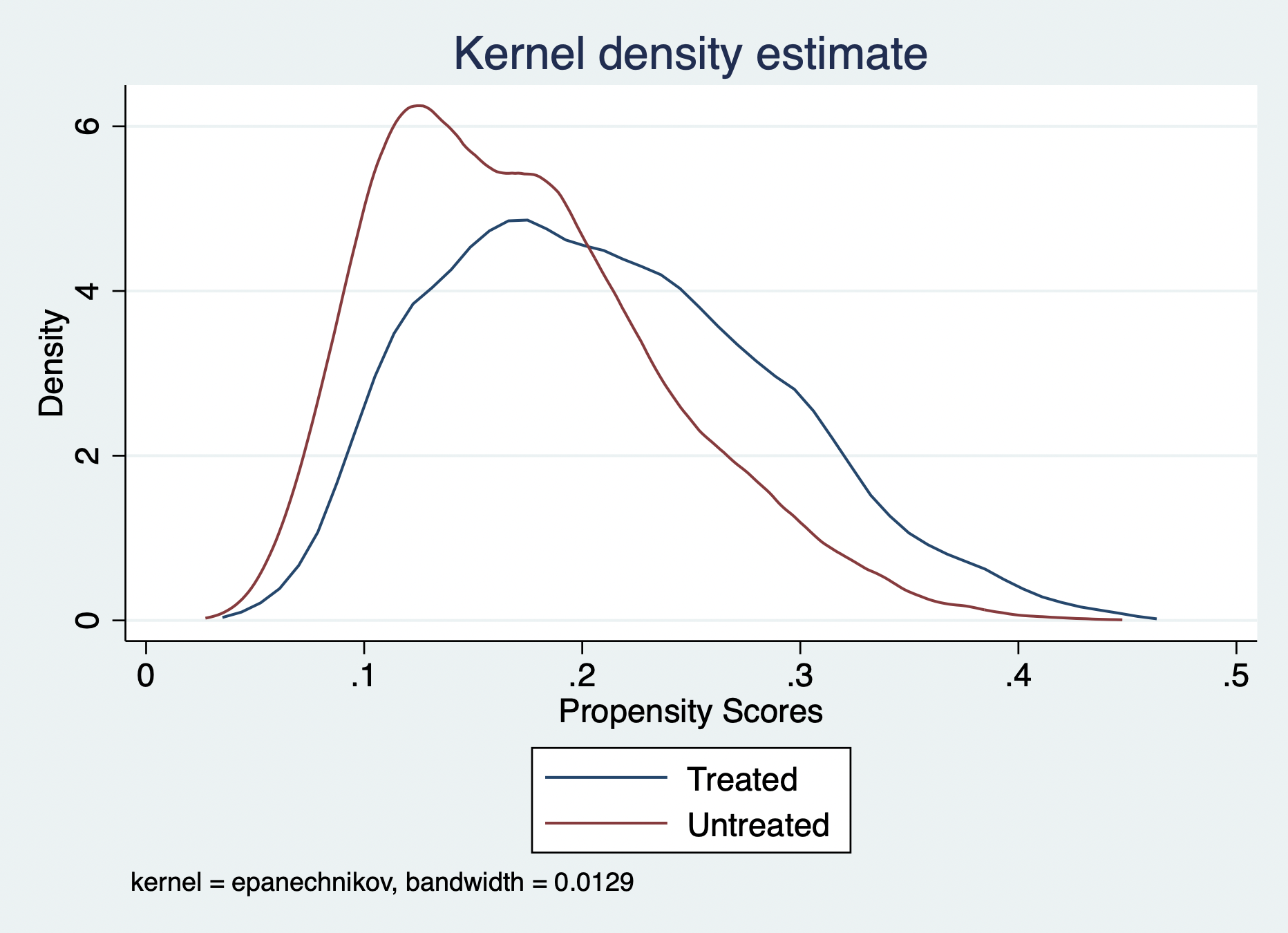}
    \caption{Propensity scores of the probability to undertake resource-saving innovations.}
    \label{fig:p_score_lme}
    \end{center}
\end{figure}

\begin{figure}[thp]
    \begin{center}
    \includegraphics[width=0.6\linewidth]{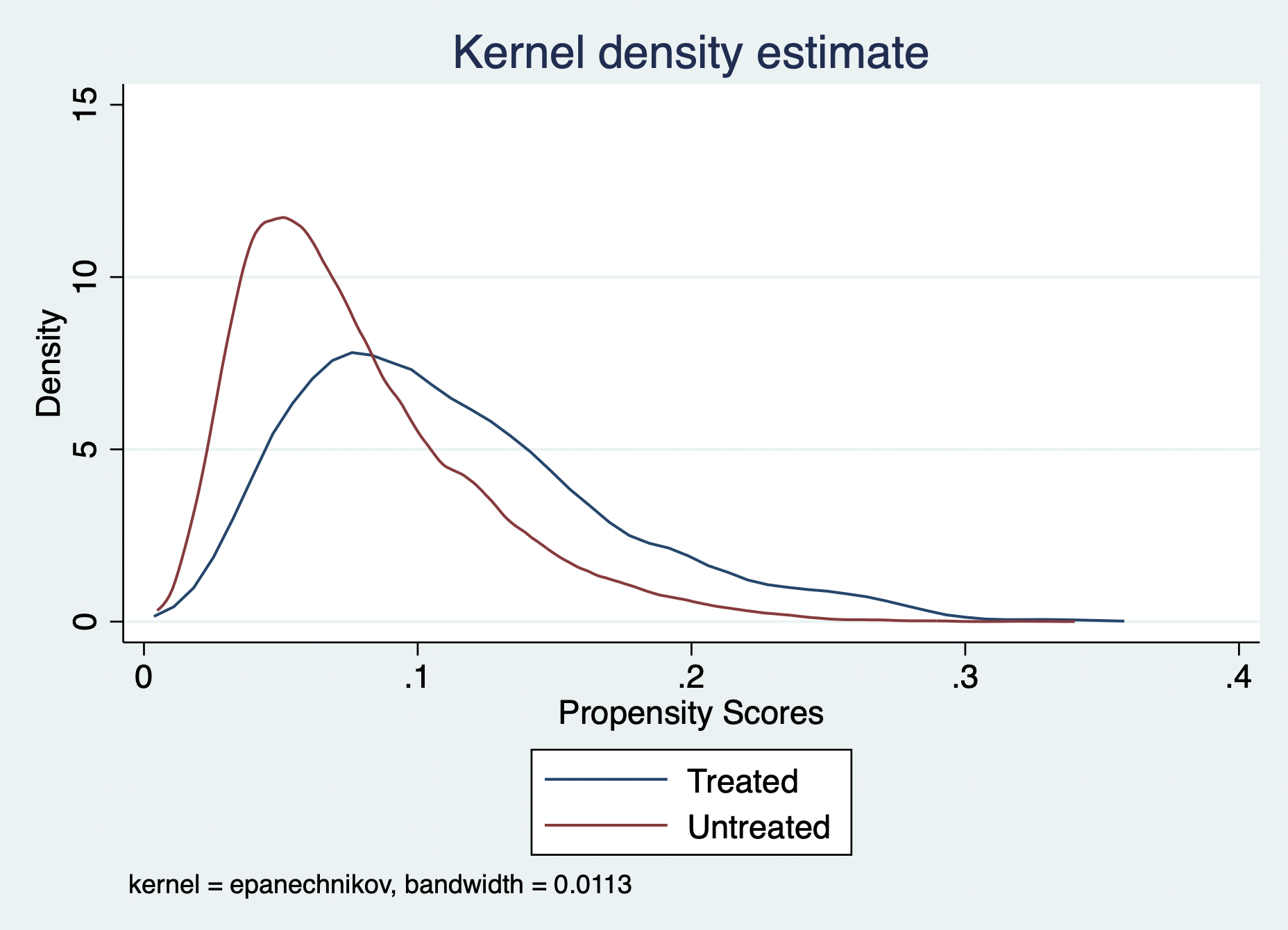}
    \caption{Propensity scores of the probability to undertake pollution-reducing innovations.}
    \label{fig:p_score_pollut}
 \end{center}
 \end{figure}

\begin{figure}[thp]
   \begin{center}
    \includegraphics[width=0.6\linewidth]{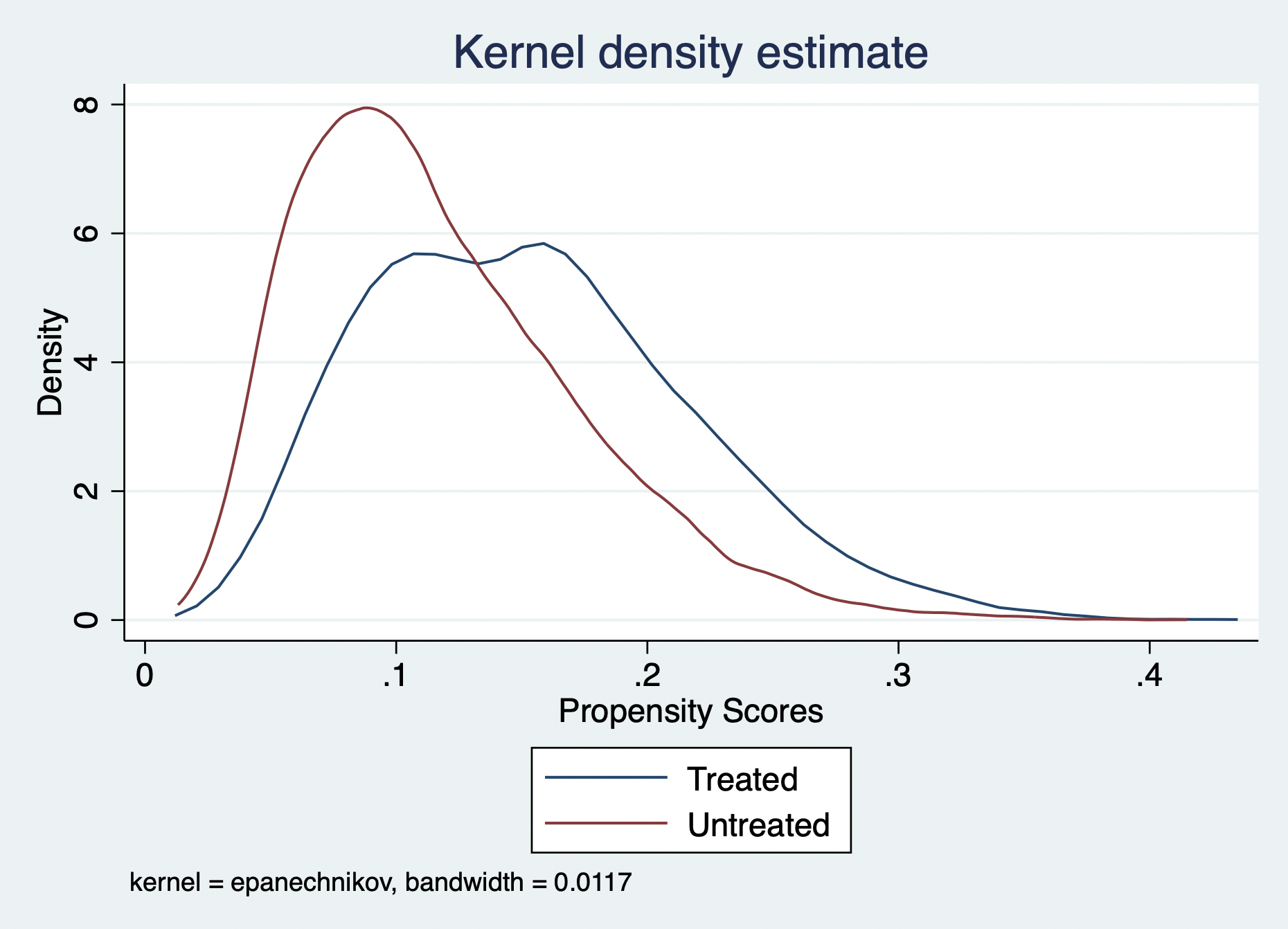}
    \caption{Propensity scores of the probability to undertake regulatory-driven innovations.}
    \label{fig:p_score_compl}
\end{center}
\end{figure}

Figures~\ref{fig:p_score_lme}, \ref{fig:p_score_pollut} and \ref{fig:p_score_compl} show the kernel density estimates of the p-scores of inclusion in either the treated or control group, respectively for each environmental innovation type. They reveal that the overlap assumption is not violated. We also observe no concentration around 0 or 1, thus essentially ruling out bias and excessive variance that may drive the estimates due to extreme p-scores.

Next, we examine standard covariate balance diagnostics (standardised differences and variance ratios across weighted vs.~unweighted treaded and control units). The results, in Tables~\ref{tab:cov-bal_resource}, \ref{tab:cov-bal_pollut} and \ref{tab:cov-bal_compl}, support that the IPWRA estimates achieve satisfactory covariate balance. In fact, the weighted standardised differences are all quite below (in absolute value) 0.25, which is the benchmark upper bound for covariates balance, while the variance ratios after weighting are all very close to the benchmark value
of~1~\citep{inbens_rubin2015, mccaffrey2004, stuart_yim2010}.

\newpage

\begin{table}[t]
    \centering 
     \caption{Covariate balance diagnostics, treated units are firms involved in resource-saving innovation}
    \label{table:covbalance_resource}
\resizebox{\textwidth}{!}{%
\begin{tabular}{@{}lllllllllllll@{}}
\toprule
                             & \multicolumn{4}{c}{Growth of Sales}          & \multicolumn{4}{c}{Growth of Employees}      & \multicolumn{4}{c}{Productivity}             \\ \midrule
 &
  \multicolumn{2}{c}{Standardized differences} &
  \multicolumn{2}{c}{Variance ratio} &
  \multicolumn{2}{c}{Standardized differences} &
  \multicolumn{2}{c}{Variance ratio} &
  \multicolumn{2}{c}{Standardized differences} &
  \multicolumn{2}{c}{Variance ratio} \\ \midrule
                             & Raw      & Weighted  & Raw       & Weighted  & Raw      & Weighted  & Raw       & Weighted  & Raw      & Weighted  & Raw       & Weighted  \\
\textit{RD\_Int\_Exp(t-1)}     & .234762  & -.002646  & .9303695  & .9999821  & .234762  & -.002646  & .9303695  & .9999821  & .234762  & -.002646  & .9303695  & .9999821  \\
\textit{Prod\_Inno(t-1)}       & .1258621 & -.0093022 & .9356835  & 1.004.029 & .1258621 & -.0093022 & .9356835  & 1.004.029 & .1258621 & -.0093022 & .9356835  & 1.004.029 \\
\textit{Proc\_Inno(t-1)}       & .2926986 & -.0149851 & .8014353  & 1.007.914 & .2926986 & -.0149851 & .8014353  & 1.007.914 & .2926986 & -.0149851 & .8014353  & 1.007.914 \\
\textit{GR\_Sales(t-1)}        & .0657836 & .0019863  & .8536271  & .9881806  & .0657836 & .0019863  & .8536271  & .9881806  & .0657836 & .0019863  & .8536271  & .9881806  \\
\textit{Employees(t-1)}        & .0142698 & -.0053808 & 1.030.868 & 1.031.418 & .0142698 & -.0053808 & 1.030.868 & 1.031.418 & .0142698 & -.0053808 & 1.030.868 & 1.031.418 \\
\textit{Age(t)}                 & .1040229 & .0084737  & .9455484  & .9959494  & .1040229 & .0084737  & .9455484  & .9959494  & .1040229 & .0084737  & .9455484  & .9959494  \\
\textit{Ext\_Fin\_Constr(t-1)} & .062701  & -.003876  & 1.101.474 & .993776   & .062701  & -.003876  & 1.101.474 & .993776   & .062701  & -.003876  & 1.101.474 & .993776   \\
\textit{Subsidy(t-1)}          & .0977485 & -.003145  & 130.838   & .9908021  & .0977485 & -.003145  & 130.838   & .9908021  & .0977485 & -.003145  & 130.838   & .9908021  \\
\textit{Patents(t-1)}          & .1063853 & -.0051879 & 1.076.433 & .995994   & .1063853 & -.0051879 & 1.076.433 & .995994   & .1063853 & -.0051879 & 1.076.433 & .995994   \\
\textit{Coop(t-1)}      & .0209451 & .0116063  & .6879709  & .69608    & .0209451 & .0116063  & .6879709  & .69608    & .0209451 & .0116063  & .6879709  & .69608    \\
\textit{Export(t-1)}           & .1839423 & -.0115332 & .8703921  & 1.007.025 & .1839423 & -.0115332 & .8703921  & 1.007.025 & .1839423 & -.0115332 & .8703921  & 1.007.025 \\
\textit{Group(t-1)}            & .1078382 & -.0163494 & 1.029.455 & .9941676  & .1078382 & -.0163494 & 1.029.455 & .9941676  & .1078382 & -.0163494 & 1.029.455 & .9941676  \\ \bottomrule
\end{tabular}%
\label{tab:cov-bal_resource}
}
\end{table}

\begin{table}[t]
     \centering 
     \caption{Covariate balance diagnostics, treated units are firms involved in pollution-reducing innovation}
\label{table:covbalance_pollut}
\resizebox{\textwidth}{!}{%
\begin{tabular}{@{}lllllllllllll@{}}
\toprule
 &
  \multicolumn{4}{c}{Growth of Sales} &
  \multicolumn{4}{c}{Growth of Employees} &
  \multicolumn{4}{c}{Productivity} \\ \midrule
 &
  \multicolumn{2}{c}{Standardized differences} &
  \multicolumn{2}{c}{Variance ratio} &
  \multicolumn{2}{c}{Standardized differences} &
  \multicolumn{2}{c}{Variance ratio} &
  \multicolumn{2}{c}{Standardized differences} &
  \multicolumn{2}{c}{Variance ratio} \\ \midrule
 &
  Raw &
  Weighted &
  Raw &
  Weighted &
  Raw &
  Weighted &
  Raw &
  Weighted &
  Raw &
  Weighted &
  Raw &
  Weighted \\
\textit{RD\_Int\_Exp(t-1)} &
  .4101037 &
  -.0057491 &
  .8813678 &
  1.028.153 &
  .4101037 &
  -.0057491 &
  .8813678 &
  1.028.153 &
  .4101037 &
  -.0057491 &
  .8813678 &
  1.028.153 \\
\textit{Prod\_Inno(t-1)} &
  .1461356 &
  .0226925 &
  .9230592 &
  .9899301 &
  .1461356 &
  .0226925 &
  .9230592 &
  .9899301 &
  .1461356 &
  .0226925 &
  .9230592 &
  .9899301 \\
\textit{Proc\_Inno(t-1)} &
  -.0220122 &
  -.0509134 &
  1.009.697 &
  1.019.494 &
  -.0220122 &
  -.0509134 &
  1.009.697 &
  1.019.494 &
  -.0220122 &
  -.0509134 &
  1.009.697 &
  1.019.494 \\
\textit{GR\_Sales(t-1)} &
  .0764506 &
  -.0328202 &
  1.078.726 &
  1.183.317 &
  .0764506 &
  -.0328202 &
  1.078.726 &
  1.183.317 &
  .0764506 &
  -.0328202 &
  1.078.726 &
  1.183.317 \\
\textit{Employees(t-1)} &
  .0999003 &
  -.0370347 &
  .9473137 &
  .9396393 &
  .0999003 &
  -.0370347 &
  .9473137 &
  .9396393 &
  .0999003 &
  -.0370347 &
  .9473137 &
  .9396393 \\
\textit{Age(t)} &
  .1082945 &
  -.0255533 &
  .9432152 &
  1.010.788 &
  .1082945 &
  -.0255533 &
  .9432152 &
  1.010.788 &
  .1082945 &
  -.0255533 &
  .9432152 &
  1.010.788 \\
\textit{Ext\_Fin\_Constr(t-1)} &
  .3153018 &
  -.0256137 &
  1.451.351 &
  .9600534 &
  .3153018 &
  -.0256137 &
  1.451.351 &
  .9600534 &
  .3153018 &
  -.0256137 &
  1.451.351 &
  .9600534 \\
\textit{Subsidy(t-1)} &
  .2454661 &
  -.0002674 &
  1.799.202 &
  .9992253 &
  .2454661 &
  -.0002674 &
  1.799.202 &
  .9992253 &
  .2454661 &
  -.0002674 &
  1.799.202 &
  .9992253 \\
\textit{Patents(t-1)} &
  .3655307 &
  -.0482171 &
  116.465 &
  .9623183 &
  .3655307 &
  -.0482171 &
  116.465 &
  .9623183 &
  .3655307 &
  -.0482171 &
  116.465 &
  .9623183 \\
\textit{Coop(t-1)} &
  .0325113 &
  .0052723 &
  .9813387 &
  1.177.553 &
  .0325113 &
  .0052723 &
  .9813387 &
  1.177.553 &
  .0325113 &
  .0052723 &
  .9813387 &
  1.177.553 \\
\textit{Export(t-1)} &
  .0986571 &
  .0142544 &
  .9374966 &
  .9916703 &
  .0986571 &
  .0142544 &
  .9374966 &
  .9916703 &
  .0986571 &
  .0142544 &
  .9374966 &
  .9916703 \\
\textit{Group(t-1)} &
  .0822034 &
  .0365699 &
  1.024.964 &
  1.011.946 &
  .0822034 &
  .0365699 &
  1.024.964 &
  1.011.946 &
  .0822034 &
  .0365699 &
  1.024.964 &
  1.011.946 \\ \bottomrule
\end{tabular}%
\label{tab:cov-bal_pollut}
}
\end{table}

\begin{table}[t]
     \centering 
     \caption{Covariate balance diagnostics, treated units are firms involved in regulatory-driven innovations}
\label{table:covbalance_compl}
\resizebox{\textwidth}{!}{%
\begin{tabular}{@{}lllllllllllll@{}}
\toprule
 &
  \multicolumn{4}{c}{Growth of Sales} &
  \multicolumn{4}{c}{Growth of Employees} &
  \multicolumn{4}{c}{Productivity} \\ \midrule
 &
  \multicolumn{2}{c}{Standardized differences} &
  \multicolumn{2}{c}{Variance ratio} &
  \multicolumn{2}{c}{Standardized differences} &
  \multicolumn{2}{c}{Variance ratio} &
  \multicolumn{2}{c}{Standardized differences} &
  \multicolumn{2}{c}{Variance ratio} \\ \midrule
\textit{RD\_Int\_Exp(t-1)} &
  .4046326 &
  -.015971 &
  .8333706 &
  1.004.549 &
  .4046326 &
  -.015971 &
  .8333706 &
  1.004.549 &
  .4046326 &
  -.015971 &
  .8333706 &
  1.004.549 \\
\textit{Prod\_Inno(t-1)} &
  .2539318 &
  -.0158528 &
  .8440063 &
  1.007.057 &
  .2539318 &
  -.0158528 &
  .8440063 &
  1.007.057 &
  .2539318 &
  -.0158528 &
  .8440063 &
  1.007.057 \\
\textit{Proc\_Inno(t-1)} &
  .0963962 &
  -.0020795 &
  .9495944 &
  1.000.959 &
  .0963962 &
  -.0020795 &
  .9495944 &
  1.000.959 &
  .0963962 &
  -.0020795 &
  .9495944 &
  1.000.959 \\
\textit{GR\_Sales(t-1)} &
  .0660297 &
  -.0201985 &
  .9091581 &
  .9599958 &
  .0660297 &
  -.0201985 &
  .9091581 &
  .9599958 &
  .0660297 &
  -.0201985 &
  .9091581 &
  .9599958 \\
\textit{Employees(t-1)} &
  .0324265 &
  .0024817 &
  1.098.704 &
  .9500144 &
  .0324265 &
  .0024817 &
  1.098.704 &
  .9500144 &
  .0324265 &
  .0024817 &
  1.098.704 &
  .9500144 \\
\textit{Age(t)} &
  .0500068 &
  .0131276 &
  .9764637 &
  .9940113 &
  .0500068 &
  .0131276 &
  .9764637 &
  .9940113 &
  .0500068 &
  .0131276 &
  .9764637 &
  .9940113 \\
\textit{Ext\_Fin\_Constr(t-1)} &
  .1412631 &
  -.0180528 &
  1.222.405 &
  .9713362 &
  .1412631 &
  -.0180528 &
  1.222.405 &
  .9713362 &
  .1412631 &
  -.0180528 &
  1.222.405 &
  .9713362 \\
\textit{Subsidy(t-1)} &
  .2225872 &
  .0059077 &
  1.722.362 &
  1.016.772 &
  .2225872 &
  .0059077 &
  1.722.362 &
  1.016.772 &
  .2225872 &
  .0059077 &
  1.722.362 &
  1.016.772 \\
\textit{Patents(t-1)} &
  .2153286 &
  .0008568 &
  113.136 &
  1.000.639 &
  .2153286 &
  .0008568 &
  113.136 &
  1.000.639 &
  .2153286 &
  .0008568 &
  113.136 &
  1.000.639 \\
\textit{Coop(t-1)} &
  .0299165 &
  .0057798 &
  1.183.017 &
  .9550862 &
  .0299165 &
  .0057798 &
  1.183.017 &
  .9550862 &
  .0299165 &
  .0057798 &
  1.183.017 &
  .9550862 \\
\textit{Export(t-1)} &
  .1043955 &
  -.0016654 &
  .933138 &
  1.000.961 &
  .1043955 &
  -.0016654 &
  .933138 &
  1.000.961 &
  .1043955 &
  -.0016654 &
  .933138 &
  1.000.961 \\
\textit{Group(t-1)} &
  -.0106958 &
  .0019833 &
  .9961558 &
  1.000.751 &
  -.0106958 &
  .0019833 &
  .9961558 &
  1.000.751 &
  -.0106958 &
  .0019833 &
  .9961558 &
  1.000.751 \\ \bottomrule
\end{tabular}%
\label{tab:cov-bal_compl}
}
\end{table}

\newpage
\clearpage
\section*{Appendix~B: IPWRA full results}

\begin{table}[thp]
\caption{IPWRA full estimates, treated units are firms involved in resource-saving innovation}
\resizebox{\textwidth}{!}{%
\begin{tabular}{@{}llllllllll@{}}
\toprule
                        & \multicolumn{3}{c}{Growth of Sales} & \multicolumn{3}{c}{Growth of Employees} & \multicolumn{3}{c}{Productivity}    \\ \cmidrule(l){2-10} 
                             & Outcome model & Outcome model & Treatment & Outcome model & Outcome model & Treatment & Outcome model & Outcome model & Treatment \\
                        & (untreated) & (treated) & model     & (untreated)  & (treated)   & model      & (untreated) & (treated) & model     \\ \midrule
\textit{RD\_Int\_Exp(t-1)}     & 0.00109       & 0.00246       & 0.0237*** & 0.000439      & 0.000882      & 0.0237*** & 0.00518***    & 0.00437       & 0.0237*** \\
\textit{}               & (0.00113)   & (0.00193) & (0.00284) & (0.000509)   & (0.000924)  & (0.00284)  & (0.00200)   & (0.00443) & (0.00284) \\
\textit{Prod\_Inno(t-1)}       & -0.0244*      & -0.0148       & 0.0866*** & -0.0139***    & -0.0119       & 0.0866*** & 0.00250       & -0.0341       & 0.0866*** \\
\textit{}               & (0.0125)    & (0.0282)  & (0.0297)  & (0.00502)    & (0.0102)    & (0.0297)   & (0.0207)    & (0.0447)  & (0.0297)  \\
\textit{Proc\_Inno(t-1)}  & -0.0132     & -0.00803  & 0.397***  & 0.00497      & 0.0149      & 0.397***   & -0.0119     & -0.0178   & 0.397***  \\
\textit{}               & (0.0107)    & (0.0245)  & (0.0306)  & (0.00469)    & (0.0105)    & (0.0306)   & (0.0194)    & (0.0463)  & (0.0306)  \\
\textit{GR\_Sales(t-1)}   & 0.0128**    & 0.0102    & 0.00154   & -0.0111***   & -0.0123*    & 0.00154    & -0.0433***  & -0.0374*  & 0.00154   \\
\textit{}               & (0.00577)   & (0.0135)  & (0.0130)  & (0.00300)    & (0.00728)   & (0.0130)   & (0.0113)    & (0.0203)  & (0.0130)  \\
\textit{Employees(t-1)}   & 0.00294     & -0.0438   & -0.0538   & 0.0234       & -0.0219     & -0.0538    & 0.195***    & 0.129     & -0.0538   \\
\textit{}               & (0.0447)    & (0.0659)  & (0.0548)  & (0.0159)     & (0.0338)    & (0.0548)   & (0.0496)    & (0.123)   & (0.0548)  \\
\textit{Age(t)}            & -0.00624    & 0.00333   & 0.116***  & -0.00808     & -0.00897    & 0.116***   & -0.141***   & -0.188*** & 0.116***  \\
\textit{}               & (0.0100)    & (0.0249)  & (0.0292)  & (0.00492)    & (0.0111)    & (0.0292)   & (0.0208)    & (0.0407)  & (0.0292)  \\
\textit{Ext\_Fin\_Constr(t-1)} & 0.0405***     & 0.0164        & -0.0373   & 0.0193***     & 0.0400***     & -0.0373   & -0.0960***    & -0.0268       & -0.0373   \\
\textit{}               & (0.0152)    & (0.0301)  & (0.0391)  & (0.00604)    & (0.0133)    & (0.0391)   & (0.0304)    & (0.0561)  & (0.0391)  \\
\textit{Subsidy(t-1)}     & -0.00393    & 0.0193    & 0.0354    & 0.0159**     & -0.00296    & 0.0354     & -0.0518     & -0.0463   & 0.0354    \\
\textit{}               & (0.0197)    & (0.0635)  & (0.0473)  & (0.00699)    & (0.0141)    & (0.0473)   & (0.0377)    & (0.0695)  & (0.0473)  \\
\textit{Patents(t-1)}     & -0.00263    & -0.00600  & 0.0380    & 0.00789      & -0.0285**   & 0.0380     & 0.0169      & 0.0649    & 0.0380    \\
\textit{}               & (0.0115)    & (0.0204)  & (0.0308)  & (0.00562)    & (0.0111)    & (0.0308)   & (0.0238)    & (0.0434)  & (0.0308)  \\
\textit{Cooperation(t-1)} & -0.282***   & -0.126**  & 0.0206    & 0.0377***    & 0.0323      & 0.0206     & 0.158***    & 0.217***  & 0.0206    \\
\textit{}               & (0.0684)    & (0.0607)  & (0.0130)  & (0.0104)     & (0.0255)    & (0.0130)   & (0.0453)    & (0.0520)  & (0.0130)  \\
\textit{Export(t-1)}      & 0.0249**    & 0.0756**  & 0.0450    & 0.0125**     & 0.0416***   & 0.0450     & 0.325***    & 0.366***  & 0.0450    \\
\textit{}               & (0.0121)    & (0.0334)  & (0.0347)  & (0.00596)    & (0.0125)    & (0.0347)   & (0.0269)    & (0.0572)  & (0.0347)  \\
\textit{Group(t-1)}       & -0.0210*    & -0.0161   & 0.0967*** & 0.00113      & -0.0150     & 0.0967***  & 0.405***    & 0.413***  & 0.0967*** \\
                        & (0.0121)    & (0.0289)  & (0.0351)  & (0.00580)    & (0.0113)    & (0.0351)   & (0.0277)    & (0.0508)  & (0.0351)  \\
Constant                & -0.129      & -0.268    & -1.545*** & -0.231**     & -0.0168     & -1.545***  & 10.47***    & 10.47***  & -1.545*** \\
                        & (0.160)     & (0.343)   & (0.212)   & (0.0917)     & (0.252)     & (0.212)    & (0.173)     & (0.364)   & (0.212)   \\
TIME FE                 & YES         & YES       & YES       & YES          & YES         & YES        & YES         & YES       & YES       \\
INDUSTRY FE             & YES         & YES       & YES       & YES          & YES         & YES        & YES         & YES       & YES       \\
Observations            & 23,128      & 23,128    & 23,128    & 23,128       & 23,128      & 23,128     & 23,128      & 23,128    & 23,128    \\ \bottomrule
\end{tabular}%
}
    \begin{tablenotes}
      \small 
    \item {\it Notes:} Robust standard errors in
      parenthesis. Asterisks denote significance levels:
      \sym{*}\(p<0.1\), \sym{**}\(p<0.05\), \sym{***}\(p<0.01\).
    \end{tablenotes}
\end{table}

\begin{table}[thp]
\caption{IPWRA full estimates, treated units are firms involved in pollution-reducing innovation}
\resizebox{\textwidth}{!}{%

\begin{tabular}{@{}llllllllll@{}}
\toprule
                        & \multicolumn{3}{c}{Growth of Sales} & \multicolumn{3}{c}{Growth of Employees} & \multicolumn{3}{c}{Productivity}    \\ \cmidrule(l){2-10} 
                             & Outcome model & Outcome model & Treatment & Outcome model & Outcome model & Treatment & Outcome model & Outcome model & Treatment \\
                        & (untreated) & (treated) & model     & (untreated)   & (treated)  & model      & (untreated) & (treated) & model     \\ \midrule
\textit{RD\_Int\_Exp(t-1)}     & 0.00126       & 0.00133       & 0.0227*** & 0.000564      & 0.000884      & 0.0227*** & 0.00518***    & 0.00437       & 0.0227*** \\
\textit{}               & (0.00113)   & (0.00682) & (0.00408) & (0.000511)    & (0.00160)  & (0.00408)  & (0.00200)   & (0.00443) & (0.00408) \\
\textit{Prod\_Inno(t-1)}  & -0.0242*    & 0.0277    & 0.0998**  & -0.0135***    & -0.0129    & 0.0998**   & 0.00250     & -0.0341   & 0.0998**  \\
\textit{}               & (0.0128)    & (0.0651)  & (0.0394)  & (0.00512)     & (0.0162)   & (0.0394)   & (0.0207)    & (0.0447)  & (0.0394)  \\
\textit{Proc\_Inno(t-1)}  & -0.0132     & 0.0226    & 0.00149   & 0.00546       & 0.0166     & 0.00149    & -0.0119     & -0.0178   & 0.00149   \\
\textit{}               & (0.0110)    & (0.0435)  & (0.0384)  & (0.00477)     & (0.0178)   & (0.0384)   & (0.0194)    & (0.0463)  & (0.0384)  \\
\textit{GR\_Sales(t-1)}   & 0.0134**    & 0.0752**  & 0.0164    & -0.0109***    & -0.00821   & 0.0164     & -0.0433***  & -0.0374*  & 0.0164    \\
\textit{}               & (0.00558)   & (0.0334)  & (0.0187)  & (0.00305)     & (0.00768)  & (0.0187)   & (0.0113)    & (0.0203)  & (0.0187)  \\
\textit{Employees(t-1)}   & -0.00309    & 0.119*    & 0.151*    & 0.0239        & 0.0162     & 0.151*     & 0.195***    & 0.129     & 0.151*    \\
\textit{}               & (0.0404)    & (0.0703)  & (0.0796)  & (0.0161)      & (0.0368)   & (0.0796)   & (0.0496)    & (0.123)   & (0.0796)  \\
\textit{Age(t)}            & -0.00722    & 0.171**   & 0.0453    & -0.00805      & -0.0280*   & 0.0453     & -0.141***   & -0.188*** & 0.0453    \\
\textit{}               & (0.0102)    & (0.0875)  & (0.0428)  & (0.00499)     & (0.0162)   & (0.0428)   & (0.0208)    & (0.0407)  & (0.0428)  \\
\textit{Ext\_Fin\_Constr(t-1)} & 0.0379**      & 0.0322        & 0.161***  & 0.0189***     & 0.000903      & 0.161***  & -0.0960***    & -0.0268       & 0.161***  \\
\textit{}               & (0.0155)    & (0.0547)  & (0.0485)  & (0.00612)     & (0.0172)   & (0.0485)   & (0.0304)    & (0.0561)  & (0.0485)  \\
\textit{Subsidy(t-1)}     & -0.00510    & -0.0194   & 0.210***  & 0.0159**      & -0.0224    & 0.210***   & -0.0518     & -0.0463   & 0.210***  \\
\textit{}               & (0.0200)    & (0.0489)  & (0.0602)  & (0.00691)     & (0.0285)   & (0.0602)   & (0.0377)    & (0.0695)  & (0.0602)  \\
\textit{Patents(t-1)}     & -0.000998   & -0.0118   & 0.240***  & 0.00666       & 0.0194     & 0.240***   & 0.0169      & 0.0649    & 0.240***  \\
\textit{}               & (0.0118)    & (0.0344)  & (0.0411)  & (0.00571)     & (0.0143)   & (0.0411)   & (0.0238)    & (0.0434)  & (0.0411)  \\
\textit{Cooperation(t-1)} & -0.285***   & -0.135*   & 0.0206    & 0.0359***     & 0.0667**   & 0.0206     & 0.158***    & 0.217***  & 0.0206    \\
\textit{}               & (0.0708)    & (0.0692)  & (0.0202)  & (0.0103)      & (0.0310)   & (0.0202)   & (0.0453)    & (0.0520)  & (0.0202)  \\
\textit{Export(t-1)}      & 0.0213*     & 0.182**   & -0.0425   & 0.0119**      & -0.0183    & -0.0425    & 0.325***    & 0.366***  & -0.0425   \\
\textit{}               & (0.0122)    & (0.0871)  & (0.0467)  & (0.00600)     & (0.0185)   & (0.0467)   & (0.0269)    & (0.0572)  & (0.0467)  \\
\textit{Group(t-1)}       & -0.0215*    & -0.0790   & 0.0271    & 0.000698      & 0.0280     & 0.0271     & 0.405***    & 0.413***  & 0.0271    \\
                        & (0.0124)    & (0.0825)  & (0.0474)  & (0.00592)     & (0.0200)   & (0.0474)   & (0.0277)    & (0.0508)  & (0.0474)  \\
Constant                & -0.123      & -0.747**  & -2.198*** & -0.234***     & 0.0401     & -2.198***  & 10.47***    & 10.47***  & -2.198*** \\
                        & (0.152)     & (0.359)   & (0.267)   & (0.0908)      & (0.105)    & (0.267)    & (0.173)     & (0.364)   & (0.267)   \\
TIME FE                 & YES         & YES       & YES       & YES           & YES        & YES        & YES         & YES       & YES       \\
INDUSTRY FE             & YES         & YES       & YES       & YES           & YES        & YES        & YES         & YES       & YES       \\
Observations            & 20,67       & 20,67     & 20,67     & 20,67         & 20,67      & 20,67      & 20,67       & 20,67     & 20,67     \\ \bottomrule
\end{tabular}%
}
    \begin{tablenotes}
      \small 
    \item {\it Notes:} Robust standard errors in
      parenthesis. Asterisks denote significance levels:
      \sym{*}\(p<0.1\), \sym{**}\(p<0.05\), \sym{***}\(p<0.01\).
    \end{tablenotes}

\end{table}

\begin{table}[thp]
\caption{IPWRA full estimates, treated units are firms involved in regulation-driven innovation}
\resizebox{\textwidth}{!}{%
\begin{tabular}{@{}llllllllll@{}}
\toprule
                       & \multicolumn{3}{c}{Growth of Sales} & \multicolumn{3}{c}{Growth of Employees} & \multicolumn{3}{c}{Productivity}    \\ \cmidrule(l){2-10} 
                             & Outcome model & Outcome model & Treatment & Outcome model & Outcome model & Treatment & Outcome model & Outcome model & Treatment \\
                       & (untreated) & (treated) & model     & (untreated)   & (treated)  & model      & (untreated) & (treated) & model     \\ \midrule
\textit{RD\_Int\_Exp(t-1)}     & 0.00101       & 0.00291       & 0.0307*** & 0.000521      & 0.000446      & 0.0307*** & 0.00502**     & 0.0103**      & 0.0307*** \\
\textit{}                    & (0.00115)     & (0.00300)     & (0.00316) & (0.000522)    & (0.00101)     & (0.00316) & (0.00201)     & (0.00490)     & (0.00316) \\
\textit{Prod\_Inno(t-1)}       & -0.0251**     & 0.0535        & 0.175***  & -0.0140***    & -0.00752      & 0.175***  & 0.00446       & 0.0442        & 0.175***  \\
\textit{}              & (0.0127)    & (0.0384)  & (0.0352)  & (0.00517)     & (0.0115)   & (0.0352)   & (0.0210)    & (0.0557)  & (0.0352)  \\
\textit{Proc\_Inno(t-1)} & -0.0143     & -0.000687 & 0.150***  & 0.00508       & 0.0147     & 0.150***   & -0.0142     & 0.0521    & 0.150***  \\
\textit{}              & (0.0108)    & (0.0198)  & (0.0334)  & (0.00473)     & (0.00900)  & (0.0334)   & (0.0196)    & (0.0474)  & (0.0334)  \\
\textit{GR\_Sales(t-1)}        & 0.0128**      & -0.0135       & 0.0310**  & -0.0115***    & -0.00742      & 0.0310**  & -0.0435***    & -0.0506       & 0.0310**  \\
\textit{}              & (0.00586)   & (0.0225)  & (0.0149)  & (0.00320)     & (0.00750)  & (0.0149)   & (0.0113)    & (0.0323)  & (0.0149)  \\
\textit{Employees(t-1)}  & 0.00171     & 0.107     & -0.0148   & 0.0245        & -0.000192  & -0.0148    & 0.202***    & 0.196     & -0.0148   \\
\textit{}              & (0.0431)    & (0.134)   & (0.0657)  & (0.0170)      & (0.0221)   & (0.0657)   & (0.0473)    & (0.135)   & (0.0657)  \\
\textit{Age(t)}           & -0.00523    & -0.00306  & 0.0332    & -0.00736      & -0.00749   & 0.0332     & -0.142***   & -0.148*** & 0.0332    \\
\textit{}              & (0.0102)    & (0.0208)  & (0.0354)  & (0.00507)     & (0.00997)  & (0.0354)   & (0.0209)    & (0.0454)  & (0.0354)  \\
\textit{Ext\_Fin\_Constr(t-1)} & 0.0413***     & 0.0460        & -0.0798*  & 0.0194***     & 0.00641       & -0.0798*  & -0.0936***    & -0.147**      & -0.0798*  \\
\textit{}              & (0.0154)    & (0.0328)  & (0.0430)  & (0.00613)     & (0.0132)   & (0.0430)   & (0.0310)    & (0.0608)  & (0.0430)  \\
\textit{Subsidy(t-1)}    & -0.00268    & -0.0569   & 0.187***  & 0.0162**      & 0.00774    & 0.187***   & -0.0520     & -0.0404   & 0.187***  \\
\textit{}              & (0.0197)    & (0.0370)  & (0.0554)  & (0.00697)     & (0.0142)   & (0.0554)   & (0.0383)    & (0.0658)  & (0.0554)  \\
\textit{Patents(t-1)}    & -0.000350   & -0.0482   & 0.140***  & 0.00739       & -0.00475   & 0.140***   & 0.0183      & 0.0131    & 0.140***  \\
\textit{}              & (0.0118)    & (0.0348)  & (0.0352)  & (0.00582)     & (0.0124)   & (0.0352)   & (0.0240)    & (0.0524)  & (0.0352)  \\
\textit{Cooperation(t-1)}      & -0.284***     & -0.198**      & 0.00527   & 0.0377***     & 0.00650       & 0.00527   & 0.165***      & 0.305***      & 0.00527   \\
\textit{}              & (0.0678)    & (0.0785)  & (0.0148)  & (0.0110)      & (0.00841)  & (0.0148)   & (0.0454)    & (0.0888)  & (0.0148)  \\
\textit{Export(t-1)}     & 0.0236*     & 8.83e-05  & -0.0297   & 0.0125**      & 0.00791    & -0.0297    & 0.323***    & 0.290***  & -0.0297   \\
\textit{}              & (0.0121)    & (0.0335)  & (0.0393)  & (0.00605)     & (0.0129)   & (0.0393)   & (0.0269)    & (0.0733)  & (0.0393)  \\
\textit{Group(t-1)}      & -0.0217*    & 0.0150    & -0.0649   & 0.000613      & -0.0100    & -0.0649    & 0.404***    & 0.430***  & -0.0649   \\
                       & (0.0123)    & (0.0215)  & (0.0405)  & (0.00595)     & (0.0132)   & (0.0405)   & (0.0279)    & (0.0556)  & (0.0405)  \\
Constant               & -0.139      & -0.300    & -1.694*** & -0.245**      & -0.0449    & -1.694***  & 10.46***    & 10.38***  & -1.694*** \\
                       & (0.161)     & (0.359)   & (0.234)   & (0.0963)      & (0.162)    & (0.234)    & (0.170)     & (0.434)   & (0.234)   \\
TIME FE                & YES         & YES       & YES       & YES           & YES        & YES        & YES         & YES       & YES       \\
INDUSTRY FE            & YES         & YES       & YES       & YES           & YES        & YES        & YES         & YES       & YES       \\
Observations           & 21,689      & 21,689    & 21,689    & 21,689        & 21,689     & 21,689     & 21,689      & 21,689    & 21,689    \\ \bottomrule
\end{tabular}%
}
    \begin{tablenotes}
      \small 
    \item {\it Notes:} Robust standard errors in
      parenthesis. Asterisks denote significance levels:
      \sym{*}\(p<0.1\), \sym{**}\(p<0.05\), \sym{***}\(p<0.01\).
    \end{tablenotes}

\end{table}

\end{document}